\documentstyle[gcdv,psfig]{article}
\def\reference{\parskip 0pt\par\noindent\hangindent 0.5 truecm}

\begin{document}

\small
\shorttitle{Star clusters}
\shortauthor{K.\ Bekki}

%
%
\title
{\large \bf
Star clusters and galactic chemodynamics: Implosive formation of super star clusters.
}
%


\author{\small 
 Kenji Bekki $^{1}$ 
} 

\date{}
\twocolumn[
\maketitle
\vspace{-20pt}
\small
{\center
School of physics,
University of New South Wales, Sydney 2052,
Australia\\
$^1$bekki@bat.phys.unsw.edu.au\\[3mm]
}

%
\begin{center}
{\bfseries Abstract}
\end{center}
\begin{quotation}
\begin{small}
\vspace{-5pt}
We numerically investigate dynamical and chemical properties of
star clusters (open and globular clusters, and ``super star clusters'', SSC)
formed in interacting/merging galaxies. The investigation is two-fold:
(1) large-scale (100pc-100kpc) SPH simulations on density and temperature evolution
of gas in interacting/merging galaxies and (2) small-scale ones on
the effects of high gas pressure of ISM on the evolution of molecular clouds.
We find that the pressure of ISM in merging galaxies can become higher than
the internal pressure of GMCs ($\sim$  $10^5$ $\rm k_{\rm B}$ K $\rm cm^{-3}$), 
in particular, in
the tidal tails or the central regions of mergers. We also find that
GMCs can collapse to form super star clusters within an order of $10^7$ yr
owing to the strong compression
by the high presser ISM in mergers.
{\bf Keywords: Galaxy: kinematics and dynamics
}
\end{small}
\end{quotation}
]


\bigskip

\section{Introduction}

Star clusters (hereafter SC) such as open, globular, and super-star clusters are
observed to be relatively ``minor'' components (in mass) in luminous galaxies
(e.g., the mass ratio of the Galactic GCs to the Galactic luminous disk mass
is an order of 0.1 \%). Therefore, it is generally considered that 
dynamical evolution of star clusters (e.g., merging or tidal destruction
of star clusters) does not influence significantly
the evolution of luminous galaxies.  However, recent numerical simulations
on dynamical evolution of SCs  have strongly suggested  
that this SC dynamics is very important for the formation of
galactic nuclei in dwarf irregulars/ellipticals
and that of ultra-compact dwarf galaxies (Drinkwater et al. 2003),
because (1) time scale of dynamical friction of SCs is relatively short 
($\sim$ $10^9$ yr) and (2) cross section of SC merging is also large owing
to the larger $\rm S_{\rm N}$ of dwarf galaxies (Bekki et al. 2003).  
Furthermore recent HST observations have revealed a significant number of 
young super star clusters (hereafter SSC) in dwarf 
irregular galaxies and suggested that 
physical properties of these SSCs can be correlated with
those of their host galaxies  (e.g., Billett et al. 2002).
These recent numerical and observational results lead the author to 
investigate (1) how SSCs can be formed in galaxies, (2) how these SSCs dynamically
evolve, and (3) whether or not there can be strong physical relationship between
structural and chemical properties of SSCs and those of their host galaxies.
In this paper,  we discuss these problems based on numerical simulations of
SSC formation in galaxies. We here focus on SSC formation in major mergers,
because recent HST observations discovered many SSCs in interacting/merging
galaxies (e.g., Whitmore et al. 1993).

\section{A new two-fold numerical model}

Since it is very hard to resolve SSC formation sites (less than pc) and
investigate the formation processes in detail {\bf in large-scale (10-100kpc) galaxy
simulations}, we adopt the following  novel two-fold method for the  numerical
investigation of SSC formation. Firstly, we investigate time evolution of pressure,
temperature, and density of warm interstellar medium (ISM) with the initial
temperature of $\sim 10^4$ K and the density of 1 atom $\rm cm^{-3}$, based
on galaxy-scale (larger-scale) SPH simulations on galaxy mergers
with total particle number ($N$) of $\sim 10^5$.
Secondly, we investigate dynamical evolution and star formation histories of
GMCs embedded by hot (warm), high pressure ISM, based on GMC-scale ($<$ 100pc)
SPH simulations ($N \sim 4 \times 10^4$). 
Primary focus here is to investigate how the high pressure ISM
(expected for galaxy mergers) trigger collapse of GMCs and  consequently
form SSCs within the GMCs. In the second simulations, the data of the ISM
surrounding GMCs come from the first simulations: The present simulations are
fully self-consistent and mimic (or corresponds to)  
the simulations with $N \sim 10^{9}$.

\subsection{Galaxy-scale}

Our numerical methods for modeling chemodynamical
evolution of galaxy mergers
have already been described  by Bekki \& Couch (2001) 
and  the details of the adopted TREESPH
are given by Bekki (1995).
We accordingly give only a brief
review of them here.
We construct models of galaxy mergers between gas-rich 
disks by using the model of Fall-Efstathiou (1980).
The total mass and the size of a progenitor exponential disk are 
$M_{\rm d}$ and $R_{\rm d}$, respectively. From now on, all the masses and 
lengths are measured in units of $M_{\rm d}$ and  $R_{\rm d}$,
respectively, unless otherwise specified. Velocity and time are 
measured in units of $v$ = $ (GM_{\rm d}/R_{\rm d})^{1/2}$ and
$t_{\rm dyn}$ = $(R_{\rm d}^{3}/GM_{\rm d})^{1/2}$, respectively,
where $G$ is the gravitational constant and assumed to be 1.0
in the present study. If we adopt $M_{\rm d}$ = 6.0 $\times$ $10^{10}$ $
\rm M_{\odot}$ and $R_{\rm d}$ = 17.5 kpc as  fiducial values, then $v$ =
1.21 $\times$ $10^{2}$ km $\rm s^{-1}$   and  $t_{\rm dyn}$ = 1.41 $\times$ $10^{8}$
yr, respectively. The dark-to-disk halo mass ratio and the star-to-gas
mass ratio are  set equal to
4.0 and 9.0, respectively.
Bulge component is not included in the present study.
An isothermal equation of state is used for the gas 
with a temperature of $7.3\times 10^3$ K (corresponding to a sound speed 
of 10 km $\rm s^{-1}$).
We present the results mostly  for  merger models with 
nearly prograde-prograde orbital configurations,  parabolic encounters,
and a  pericentric distance of 0.5 $R_{\rm d}$. 

\subsection{GMC-scale}

We numerically investigate the hydrodynamical effects of the high pressure ISM
on a self-gravitating molecular gas cloud in major mergers.
The  cloud is represented by 20,000 SPH particles and
the initial cloud mass ($M_{\rm cl}$) and  size ($r_{\rm cl}$)
are set to be $10^{6}\,M_{\odot}$ and 97\,pc, respectively,
which are consistent
with the mass-size relation observed by Larson (1981).
The  cloud is assumed to have an isothermal radial density profile
with $\rho (r) \propto 1/(r+a)^2$, where $a$ is the core radius of the cloud
and set to be $0.2r_{\rm cl}$.
An isothermal equation of state with a sound speed of $c_{\rm s}$
is used for the gas, and $c_{\rm s}$ is set to be 4\,km\,s$^{-1}$
(consistent with the prediction from the virial theorem)
for models with $M_{\rm cl} = 10^{6}\,M_{\odot}$.
Each SPH particle is also subject to the hydrodynamical force of the ISM whose
strength depends on the parameters of the ISM. The ISM is represented by 
$\sim$ 20000 SPH particles and 
and an isothermal equation of state with pressure $p$ (or sound velocity
of $c_{\rm h}$),  and density, $\rho$. 
The ISM particles are uniformly distributed within  a box  with size 
of  $6R_{\rm cl}$.
A gas particle in a cloud is converted into a collisionless stellar particle 
if (1)\,the local dynamical time scale [corresponding to  
${(4 \pi G\rho_{i})}^{-0.5}$, where $G$ and $\rho_{i}$ are the gravitational
constant and the density of the gas particle, respectively]
is shorter than the sound crossing time (corresponding to $h_{i}/c_{\rm s}$, 
where $h_{i}$ is the smoothing length of the gas), 
and (2)\,the gaseous flow is converging.
This method thus mimics star formation due to Jeans instability in gas clouds. 
In the model without  ISM effects, star formation does not occur at all in this
star formation method.
We mainly  show the result of the ``fiducial model''
with$\rho$ = 1 atom cm$^{-3}$ and $P$ = $10^5$ $\rm k_{\rm B}$ K cm$^{-3}$, 
because this model shows the typical behavior of ISM-induced star formation
in a gas cloud.

\begin{figure}
 \begin{center}
 \begin{tabular}{cc}
 \psfig{file=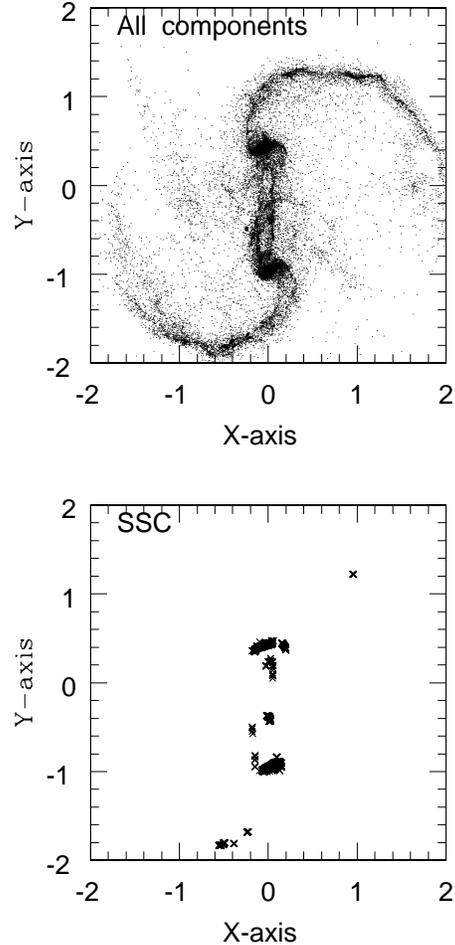,width=15.5cm} 
 \end{tabular}
\caption{
Mass distribution projected onto the $x$-$y$ plane in  the merger model
with   at $T$ = 0.28 Gyr for all components (upper) and
for formation sites (gas or gas forming SSCs)
of SSCs  with $P$ $>$ $10^5$ $k_{\rm B}$ (lower).
In total, 2667 SSC formation sites 
with $P_{\rm g}$ $>$ $10^5$ $k_{\rm B}$ ($k_{\rm B}$ is Boltzmann's constant)
have been identified  prior to  this starburst epoch.
The scale is given in our units (17.5 kpc), and each of the frames
measures 70 kpc on a side.
}
 \label{flogo}            
 \end{center}
\end{figure}

\section{Result I. SSC formation sites.}

Figure 1 clearly demonstrates how the  gaseous pressure of the interstellar medium
becomes dramatically higher ($10^4$-$10^5$ $k_{\rm B}$)
in some regions of the major merger model at $T$ = 0.28 Gyr
(starburst phase with the star formation rate of $\sim$ 75 $M_{\odot}$ ${\rm yr}^{-1}$)
in comparison with the initial disks.
Furthermore, gaseous pressure for  some fraction of the gas particles
(in particular, in the  central regions)
exceeds the threshold
value of  $2.0\times 10^5 k_{\rm B}$ for SSC formation,  
essentially because rapid transfer
of gas to the central region of the merger and 
efficient gaseous shock dissipation increase the gaseous density and pressure greatly. 
These results clearly explain  the reason why SSCs can be formed
in starbursting galaxies  and imply that the formation efficiency of SSCs are much higher 
in merging galaxies with massive starbursts than in ``normal'' disk galaxies.
Figure 1 also shows that SSCs are formed not only in the central region,
where gas fueling is more efficient,  but also
in the outer tidal tails and ``bridges'' among the two cores,
where the tidal force of the merger forms thin and high-density 
gaseous layers  owing to gaseous shock dissipation
(More details are given in Bekki \& Couch 2001).  

\section{Result II. Implosive SSC formation.}

Figures 2 and 3 describe how a burst of star formation within a gas cloud
can be triggered by the external gas pressure of the ISM in the fiducial model.
In this model with moderately strong ISM  pressure,
the high pressure of the ISM can continue to strongly compress 
the cloud without losing a significant amount of gas from the cloud.
As the strong compression proceeds, 
the internal density/pressure of the cloud can rise significantly.
However, the self gravitational 
force of the cloud also becomes stronger
because the cloud becomes progressively more compact during the ISM's compression. 
Therefore, the internal gaseous pressure of the cloud
alone becomes unable to support itself against the combined effect of
the external pressure from the ISM and the stronger self-gravitational force.
As a result of this,
the cloud's collapse initially induced by the high external pressure from the 
ISM can continue in a runaway manner.

Due to the rapid, dissipative collapse, 
the gaseous density of the cloud dramatically rises and 
consequently star formation begins in the central regions of the cloud.
The star formation rate increase significantly 
to 1.5 $M_{\odot}$ yr$^{-1}$ (8\,Myr after the start of the cloud's collapse).
About 80 \% of the gas is converted into stars within 14\,Myr
to form a stellar system. Because of the ``implosive'' formation of stars from
strongly compressed gas, the developed stellar system is strongly
self-gravitating and compact. This result implies that high external pressure
from the ISM is likely to trigger the formation of bound, compact star
clusters rather than unbound, diffuse field stars. The final mass distribution
of this model is similar to the King profile, but too compact (with the
effective radius of $\sim$ 2 pc) to be consistent with the observed
effective radius of typical globular clusters (See Figure. 4).





\begin{figure}
 \begin{center}
 \begin{tabular}{cc}
 \psfig{file=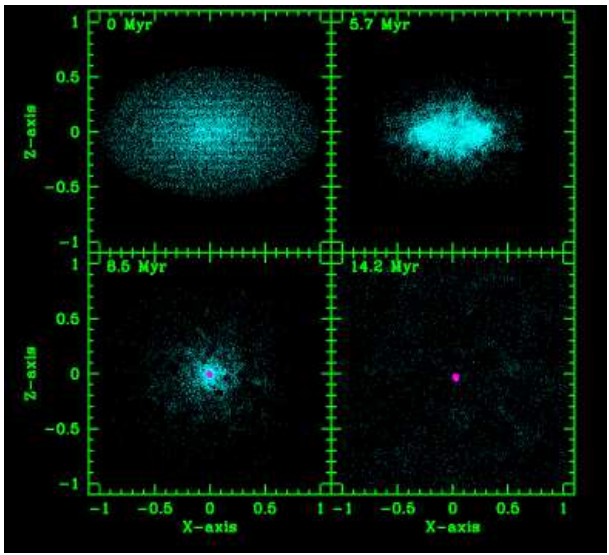,width=8.0cm} 
 \end{tabular}
\caption{
Time evolution of a GMC embedded by high pressure ISM for gaseous
components (cyan) and for new stars formed from the gas (magenta).
One frame measures 200 pc on a side. Note that a very compact SSC
is formed owing to strong external compression of the GMC by
the hot, high pressure ISM.  
}
 \label{flogo}            
 \end{center}
\end{figure}

\begin{figure}
 \begin{center}
 \begin{tabular}{cc}
 \psfig{file=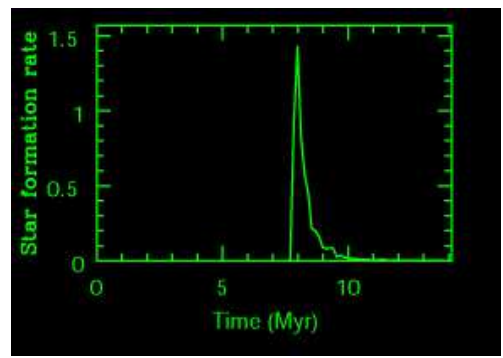,width=6.5cm} 
 \end{tabular}
\caption{
Time evolution of star formation rate of the model in Figure 2.
Note that the time scale of starburst is less than a few Myr,
which is shorter than the lifetime of very massive stars.
This implies that supernovae feedback effects are less likely
to influence  super star cluster formation.
}
 \label{flogo}            
 \end{center}
\end{figure}

\begin{figure}
 \begin{center}
 \begin{tabular}{cc}
 \psfig{file=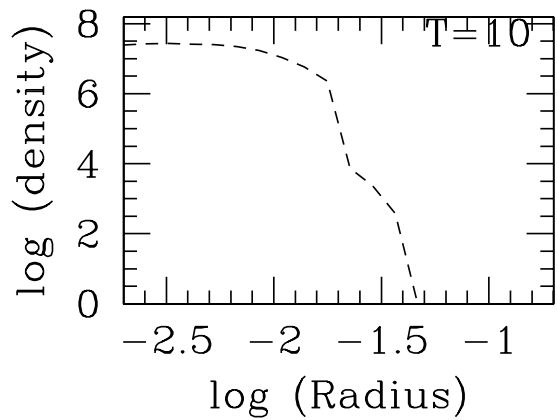,width=12.5cm} 

 \end{tabular}
\caption{
Final projected surface density distribution of a SSC developed
in the model in Figure 2 (in our units). 
Since the system has not yet relaxed dynamically just after 
the formation of the SSC, the density profile shows a hollow
around log (R) = -1.7.
}
 \label{flogo}            
 \end{center}
\end{figure}

\section{Conclusions.}

We first demonstrated that SSCs can be formed from GMCs ``in an implosive manner'' for
major mergers: Strong external pressure of ISM in mergers can trigger the 
collapse of GMCs and consequently cause the rapid star formation within
the GMCs. The star formation efficiency in the GMCs under strong external
pressure is as high as $60-80$ \% (in total cloud mass) so that the 
developed SSCs can be considered to be bound star clusters (e.g., Hills 1980).
The present study suggests that 
unbound star clusters that are  doomed to be destroyed by some external
processes (e.g., galactic tidal field and interaction with GMC) 
can be formed from GMCs embedded by  low (normal) pressure ISM  whereas
SSCs can be formed from GMCs embedded by  very high  pressure ISM.
The stars initially within 
unbound star clusters could soon be identified as  ``field stars''  after the 
destruction of the clusters. 
The time scale of SSC formation is very short (a few times 10$^6$ yr, which is
shorter than the lifetime of very massive stars, which are responsible for
the type II supernovae) so that  the energy feedback from supernovae can
be much less likely to influence the SSC formation.
Although the  ``implosive formation scenario''  described
here can provide a clue to the origin of GCs in general, we have not
yet reproduced  the observed Fundamental Plane of GCs (in M31 and the Galaxy) 
and mass-size relation of GCs.
Therefore, it is our future studies to check whether the implosive scenario
can explain self-consistently the observed scaling relations of GCs and
chemical properties of GCs.

\section*{Acknowledgments}


\section*{References}


\reference Bekki, K.  1997, ApJ, 483, 608 

\reference Bekki, K., \and Couch, W. J. 2001, 
 ApJL, 557, 19 

\reference Bekki, K., Couch, W. J.,  Drinkwater, M. J.,  \and Shioya,  Y. 2003, 
 MNRAS, in press. 

\reference Billett, O., Hunter, D. A., \and  Elmegreen, B. G., 
2002, AJ, 123, 1454


\reference  Drinkwater, M. J.,    Gregg, M. D., Hilker, M., Bekki, K., Couch,
W. J., Ferguson, H. C., Jones, J. B., \and  Phillipps, S.  2003, Nature, 423, 519


\reference Fall, S. M., \& Efstathiou, G. 1980, MNRAS, 193, 189 

\reference Hills, J. G. 1980, ApJ, 235, 986 

\reference Larson, R. B. 1981, MNRAS, 194, 809 

\reference Whitmore, B. C., Schweizer, F. Leitherer, C. 
Borne, K., \and  Robert, C.  1993, AJ, 106, 1354



\end{document}